\documentclass[12pt]{iopart}
\usepackage{graphicx}

\usepackage{iopams}  
\usepackage{caption}
\begin{document}
	
	\title[Reexamination of the $\beta^+$ decay process of $^{13}\rm{N}$ nucleus]{Reexamination of the $\beta^+$ decay process of $^{13}\rm{N}$ nucleus}
	
	\author{B. F. Irgaziev$^1$ $^2$, Jameel Un Nabi$^1$ and A.Kabir$^1$}

	\address{$^1$ GIK Institute of Engineering Sciences and Technology, Topi, Pakistan}
		\address{$^2$National University of Uzbekistan, Tashkent,Uzbekistan}
	\ead{irgaziev@yahoo.com}
	\vspace{10pt}
	\begin{indented}
		\item[]August 2019
	\end{indented}
	
	\begin{abstract}
		It is known from the scattering theory that the phase-shift of elastic collision does not provide a unique potential to describe the bound state of the two-particle system. The bound state wave function is the most crucial input for various nuclear processes bearing astrophysical significance. In this paper, we emphasize on the important role of the asymptotic normalization coefficients for description of the $^{13}\rm{N}\longrightarrow {^{13}\rm{C}}+\beta^++\nu_e$ reaction.  Using experimental data of the elastic scattering phase-shift of proton and neutron on $^{12}\rm{C}$, we find the asymptotic normalization coefficients which determine the tail of the bound state wave functions. This allows us  to modify the wave functions of $^{13}\rm{N}$ and  $^{13}\rm{C}$  within the single-particle approach using the inverse scattering theory.  These wave functions are used to determine the nuclear matrix element for calculation of the half-life  and log{\textit ft} values of $\beta^+$ decay of $^{13}\rm{N}$. The calculated values of the log{\textit ft} and half-life are smaller than the measured values when the single-particle wave functions are employed.  Using overlap functions, instead of single-particle functions, we obtain a better comparison. The overlap function is represented as the product of single-particle function and the corresponding spectroscopic factor.
	\end{abstract}

	\noindent{ Keywords}: CNO cycle, $\beta^{+} \rm decay$, potential cluster model, asymptotic normalization constant, spectroscopic factor

	\maketitle

\section{Introduction} \label{introduc}
Beta decay is one of the common modes of radioactive decay for unstable nuclei and has an important role in nuclear astrophysics. The mechanism of this process is well described in the literature (see~ \cite{Kenneth, Suhonen} and references therein). The study of beta decay provides good information for nuclear spectroscopy and astrophysics. Pauli's neutrino hypothesis plays a central role in the beta decay. Any nucleus that decay by positron emission has another competing  branch of decay, namely atomic electron capture. During the beta decay of  the nucleus, the lepton charge conservation law is fulfilled. For this reason, in addition to the daughter nucleus, we observe emission of electron (positron) and antineutrino (neutrino). The results of application of the Fermi theory \cite{Fermi1934} show that the observed distribution of energy is distributed between the daughter nucleus, beta particle, and neutrino. We may neglect the kinetic energy of the heavy daughter nucleus and assume that the energy is shared between the beta particle and neutrino \cite{Kenneth}. However, there is an inherent  problem in the calculation of the nuclear matrix element as the nuclear wave functions of parent and daughter nuclei are calculated by using different nuclear models and can lead to disparate calculated values of log\textit{ft} and  half-life for a given nucleus. To improve the result of calculated half-life value of $\beta$-decay, a phenomenological shape factor  is included to the integrand of the equation determining the rate of $\beta$ transition.  As mentioned in Ref. \cite{gross} the single-particle shell model gives \textit{ft}-values of allowed transitions generally smaller than the observed ones by factors of 10 or more if transition is carried out to several final states. One can see in Table 7.5 of Ref. \cite{Suhonen} that few calculated half-life values are less than the measured ones for certain nuclei. The gross theory \cite{gross}, the macroscopic model  based on the sum rule of the $\beta$-decay strength function, was suggested to treat the transitions to all final nuclear levels in a statistical manner. Its application is preferable for decay of heavy nuclei.  The squared matrix element of transition is calculated as an integral with respect to the energy, whose integrand is the product of some probability and the distribution function of nucleons over energy. However, for light nuclei application of the gross theory is not desirable because number of energy levels is too few to adopt a statistical approach. The shell model and the pn-QRPA model  (see references \cite{Sta90,Hir93}) are the two most frequently used microscopic models to study the beta decay of heavy nuclei in the mass range $A=40-100$. Later the models were extended to perform weak rate calculations in  stellar matter \cite{Nab99, Nab04, Lan20}. However, in the last five  references a global approach was adopted to estimate weak interaction rates for hundreds of nuclei of varying species. In Refs.  \cite{sinhgam1986, amos1989} the authors  suggested to apply the phenomenological many-particle wave functions of $^{13}$C and  $^{13}$N  to study $^{13}{\rm{C}}(\gamma, \pi^-)^{13}{\rm{N}}$ and the electromagnetic and hadronic interaction form factors for the ground, and excited states in $^{13}$C and $^{13}$N nuclei. However, in Ref. \cite{sinhgam1986} four independent parameters of the wave function were defined by fitting to the magnetic moments of $^{13}$C, $^{13}$N, the experimental elastic transfer electron scattering M1 form factor, and   log{\textit ft} of  $^{13}$N decay. The wave function of Ref. \cite{amos1989} led to value of $B_{GT}$ larger than experimental value.  Unlike the above mentioned methods, we offer a simple recipe for the calculation of  $\beta$ decay of light nucleus  when the transition from the ground state of the parental nucleus is allowed to the ground state of the daughter nucleus having pronounced single-particle configurations. We address the many-particle aspects in the weak decay of nucleus through the asymptotic normalization coefficients (ANCs) and the spectroscopic factors. 

The decay rate in the Fermi theory is subdivided according to the angular momentum of emitted leptons and there are two sorts of contribution, i.e. the Fermi and  Gamow-Teller terms \cite{GT}.

It is well known that a unique value of the ANC can be found directly from the elastic scattering phase-shift, without invoking intermediate potential, by fitting the scattering phase-shift and extrapolation the scattering amplitude from positive energy region  to the bound-state pole. Such procedure was demonstrated in Ref. \cite{blokh-93} for an $\alpha$--$d$ system. The $\alpha$--$d$ radiative capture was calculated in Ref. \cite{abi} using the ANC value extracted in Ref. \cite{blokh-93} and  ground state wave function of $^6\rm{Li}$ in the two-body approach using the inverse scattering problem theorem \cite{Chadan-77} (see also chapter 20 of the Ref. \cite{Newton}).  It should be noted that the ANC value allows one to correctly normalize the single-particle wave function. Once the Pauli principle is taken into account applying potentials with the forbidden Pauli principle bound states, it correctly describes the behavior of a single-particle wave function  in the inner region of the nucleus as well.

Beta decay of one proton  nucleus $^{13}\rm{N}$ has a very important role in the CNO-I cycle. The application of the single-particle shell model gives lower values for log\textit{ft} and half-life  and are smaller compared to the experimental results.  It is known that the $pp$ fusion chain is driving the nuclear burning in Sun but the CNO is the main burning cycle in stars heavier than the Sun where the core temperature exceeds seventeen  million degrees. It is also anticipated that some of the solar neutrino flux originates from the CNO cycle. The CNO cycles produce significant neutrino and $\beta$ particle flux due to  decays of $\rm ^{13}N$, $\rm ^{15}O$ and $\rm ^{17}F$.

In this paper we determine the log\textit{ft} and half-life values for  $^{13}\rm N$ decay by two methods: the method described in Ref. \cite{abi} and a new method using the overlap functions instead of the single-particle functions. Both methods require knowledge of the ANCs. To determine the value of the  ANC for  $^{13}\rm N$ and $^{13}\rm C$ bound state functions, fitting procedure  of  various experimental phase-shift data of the elastic scattering of nucleon on $^{12}\rm{C}$ are used.

The paper is organized as follows. In Section~\ref{beta decay} we present the the $\beta$ decay transition  in the single-particle approach. The cluster model for description of $^{13}\rm{C}$  and $^{13}\rm{N}$ nuclei is  given in Section~\ref{cluster model}. In Section~\ref{ANC determination} we present the formula for proving the effective range and $\Delta$-methods for extraction of ANCs for $^{13}\rm{C}$  and $^{13}\rm{N}$ nuclei. The results of extractions of ANCs from the  elastic scattering  phase-shifts of neutron and proton by $^{12}$C  is described in Section~\ref{extraction}. In Section~\ref{new method} we apply the new method for calculation of the $\beta$ decay of $^{13}\rm{N}$ nucleus taking into account the spectroscopic factors. We summarize our findings in Section~\ref{summary}. 

We use natural unit system ($\hbar=c=1$)  throughout this paper.

\section{The $\beta^+$ decay transition probability} \label{beta decay}
The $\rm \beta$ decay is caused by  weak interaction and the nuclear transition may be treated using the Fermi's golden rule. The transition amplitude is
\begin{equation} \label{Vif}
	M_{fi}=\langle\Psi_f|V|\Psi_i\rangle,
\end{equation}
where $V$ is a potential of weak interaction that causes the transition from an initial quantum state $\Psi_i$ (a wave function of the parent
nucleus) to a final one, $\Psi_f$ (a product of the wave function of the daughter nucleus,  positron  and  neutrino). We  ignore the neutrino rest mass.
For the treatment of  $^{13}\rm{N}$ and $^{13}\rm{C}$ nuclei, the shell model is one of the preferred models. Light nuclei have a pronounced cluster structure, and $^{13}\rm{N}$, $^{13}\rm{C}$ being mirror nuclei  are such nuclei. These nuclei can be considered as a bound state of  $^{12}\rm{C}$ nucleus and an additional proton or neutron, respectively. The probability of this configuration is higher in  $^{13}\rm{N}$ and $^{13}\rm{C}$ . The spin and isospin of $^{13}\rm{N}$ ($^{13}\rm{C}$) is defined by the total momentum and isospin of the proton (neutron) in $p_{1/2}$ level.   In the decay process of  $^{13}\rm{N}$ nucleus, only proton located at the $p_{1/2}$ external level can emit positron and neutrino, and the resulting neutron should remain at the same level. Protons located in deeper $s_{1/2}$ and $p_{3/2}$ levels cannot emit positron and neutrino, because the transformed neutron  must get transferred to the $p_{1/2}$ state due to the Pauli principle and such a transfer is not possible owing to energy conservation law. So, the wave function of the  $^{12}\rm{C}$ cluster remains unchanged. As the recoil energy during the decay of the $^{13}\rm{N}$ nucleus may be neglected, we can assume that the center of mass of the nucleus remains fixed. It is also permissible to disregard the interaction of neutrinos with the daughter nucleus and the positron. We take into account the interaction of the positron with the nucleus in the relativistic assumption, as it may have kinetic energy exceeding its rest mass-energy.

The wave functions of the emitted positron and neutrino are cut owing to the bound wave function of  $^{13}\rm{C}$ nucleus
at distance approximately equal to the radius of  $^{13}\rm{C}$ nucleus.
The de Broglie wavelength of the neutrino is  essentially larger than the effective size of the $^{13}\rm{C}$ implying the plane wave is replaced by unity.  For the wave function of the positron, coordinate is taken at the surface of the $^{13}\rm{C}$ nucleus.

The resulting expression for determining the transition probability is given by
\begin{equation}\label{lambda}
	\lambda = \frac{G^2_F\overline{|{\cal{M}}_{fi}|^2}}{2\pi^3\hbar^7c^3}\int_{0}^{p_{max}}F(-Z_f,p)p^2(Q-T_{e})^2dp,
\end{equation}
where $G_F$ is the Fermi coupling constant, the over-line means averaging over initial and summation over final state quantum numbers, $F(-Z_f,p)$ is the Fermi function \cite{jackson1958}; $Q$ is Q-value of the reaction; $T_e$ is the kinetic energy of the positron; the integration is performed over the momentum of the positron;  ${\cal{M}}_{fi}$ in Eq. (\ref{lambda}) represents the nuclear matrix element which is independent of the momentum of the positron \cite{wong} and defined as
\begin{equation}
	{\cal{M}}_{fi}= \langle\psi_f| \hat O\|\psi_i\rangle=\langle\psi_f|g_{\rm V}\mathbf{1}
	{\tau}+g_{\rm A}{\sigma\tau}|\psi_i\rangle, \label{Mfi}
\end{equation}
where the first and second  term are responsible for Fermi and Gamow-Teller decays, respectively; $g_{\rm V}=1$, $g_{\rm A}=1.25$ are the vector and axial-vector coupling constants of the weak interactions; $\mathbf{1}$ is identity operator in spin space, leaving the total momentum $J$ of nucleon on $p_{1/2}$ state unchanged and ${\sigma}$ is the Pauli spin matrices, which leads to $\Delta J=0,\pm 1$; ${\tau}$ is isospin transition matrix which turns proton to neutron; $\psi_f$ is the single-particle $^{12}\rm{C}$-$n$ wave function; $\psi_i$ is the single-particle $^{12}\rm{C}$-$p$ wave function.
In deriving Eq. (\ref{Mfi}), we assumed that the operator $\hat O$ acts only on a proton in the $p_{1/2}$ state.
We denote the integral contained in Eq. (\ref{lambda}) as
\begin{equation}\label{phase-space}
	f_0^{(+)}=\frac{1}{(m_{\rm e}c)^3(m_{\rm e}c^2)^2}\int_{0}^{p_{max}}F(-Z_f,p)p^2(Q-T_e)^2dp,
\end{equation}
where the constants have been included to make $f_0^{(+)}$ dimensionless.
The Fermi function $F(-Z_f,p)$ equals to
\begin{equation}\label{FZ-f}
	F(-Z_f,p) =  \frac{2(1 + s)}{\Gamma(1+2s)^2}(2p\rho)^{2s - 2}{\rm e}^{\pi\eta}|\Gamma(s + \rm i\eta)|^2,
\end{equation}
where  $s=(1-Z^2_f\alpha^2)^{1/2}$; $\alpha$ is the fine-structure constant; $\rho=R_{\rm N}/\hbar$ ($R_{\rm N}$ is a radius of $^{13}\rm{N}$ nucleus); $\eta=\frac{-Z_f\alpha}{v/c}$ is the Sommerfeld parameter, $v/c$ is ratio of the positron velocity to light speed.

The function $f_0^{(+)}$ is called the phase-space factor for $\beta$ decay. Experiments related to the study of the $\beta^+$ decay are performed with unstable atoms, so we must add to the factor $f_0^{(+)}$ an additional factor $f_0^{(EC)}$ that takes into account the capture of an electron from the $K$ and $L$  shells of the atom. Note, the factor $f_0^{(EC)}$ is much less ($\sim 0.2\%$) compared to $f_0^{(+)}$ in the case of the $^{13}\rm{N}$ decay.
The  $^{13}\rm{N}$ decay of probability  may be written as
\begin{equation} \label{total-lambda}
	\lambda=\frac{G^2_F\overline{|{\cal{M}}_{fi}|^2}m_{\rm e}^5c^4}{2\pi^3\hbar^7}f_0,
\end{equation}
where
\begin{equation}\label{f0}
	f_0=f_0^{(+)}+f_0^{(EC)}.
\end{equation}
The transition probability is related to the half-life by the relation
\begin{equation}\label{lambda-T}
	\lambda = \frac{\ln2}{T_{1/2}}.
\end{equation}
Using Eq. (\ref{total-lambda}) and Eq. (\ref{lambda-T}) we can write the expression for the half-life as
\begin{equation}\label{T1-2}
	T_{1/2}=\frac{2\pi^3\hbar^7\ln2}{f_0 \overline{|{\cal{M}}_{fi}|^2}m_{\rm e}^5c^4G_F^2},
\end{equation}
and substituting
\begin{equation}\label{Mifav}
	\overline{|{\cal{M}}_{fi}|^2}={\langle\varphi_f|\varphi_i\rangle}^2(g_{\rm V}^2+g_{\rm A}^2/3),
\end{equation}
calculated in the single-particle approach we obtain for $^{13}\rm{N}$ decay 
\begin{equation}\label{half-life}
	T_{1/2}=\frac{\kappa}{f_0 \langle\varphi_f|\varphi_i\rangle^2(g_{\rm V}^2+g_{\rm A}^2/3)},
\end{equation}
where ${\langle\varphi_f|\varphi_i\rangle}$ is the overlap integral of the initial and final radial wave functions and $\kappa$ is an expression consisting of constants:
$$\kappa =\frac{2\pi^3\hbar^7\ln2}{m_{\rm e}^5c^4 G_F^2}.$$
Note, this overlap integral in the single-particle shell model equals  unity. In reality, it should be less than one.
Writing ${\rm log}ft$  for $^{13}\rm{N}$ decay as
\begin{equation}\label{logft}
	{\rm log}ft={\rm log}\frac{\kappa}{\langle\varphi_f|\varphi_i\rangle^2(g_{\rm V}^2+g_{\rm A}^2/3)},
\end{equation}
we can see that ${\rm log}ft$ depends on the initial and final wave functions (of parent and daughter nuclei, respectively)   and does not depend on the emitted positron and neutrino states.

Knowing the experimentally measured half-life value for $\beta$ decay, it is possible to estimate the  ${\rm log}ft$ value using the phase-space factor [Eq. (\ref{f0})] in the case of $\beta^+$ decay or in the case of $\beta^-$ decay replacing $-Z_f$ by $Z_f$ in  Eq. (\ref{phase-space}). We have done such an estimation for the light nuclei and obtained excellent agreement with the experimental results for ${\rm log}ft$ presented in Ref. \cite{nucl-data}, including  nuclei with longer half-lives, e.g. $^{10}\rm{Be}$ and $^{14}\rm{C}$. The solution to the problem of calculating  ${\rm log}ft$ and half-life values is directly linked with the accurate computation of  associated nuclear matrix elements  ${\cal{M}}_{fi}$ in Eq.~(\ref{total-lambda}).

\section{The ground state wave functions of $^{13}\rm{N}$ and $^{13}\rm{C}$ in the cluster approach}\label{cluster model}
At low collision energies (below inelastic collision threshold), where only the channel of elastic scattering is open, it is quite reasonable to analyze the phase-shifts of  scattering within the framework of the-two particle potential approach. As is known, the two-particle binding energy, the ANC of the wave function of the bound state  and the phase-shifts  are on-shell characteristics  and there is an unambiguous  relationship between them.  The partial amplitude of elastic scattering is an analytical function. The pole of the scattering amplitude at the negative energy  determines the binding energy while the ANC is defined through the residue of the amplitude at this pole.

The use of wave functions obtained from the very laborious solution of the many-particle Schrödinger equation does not always lead to good agreement with the available experimental data. At the same time, a simple model based on physical principles may lead to a good result in decent agreement with measured data.
The model which we consider here for the theoretical calculation of the nuclear wave function is a potential cluster model (PCM) where the nucleus is treated as two clusters ($A_1$ and $A_2$) moving relative to each other. The PCM is based on the assumption that nuclei have a cluster structure. To describe the relative motion of clusters, a phenomenological potential is defined and may contain states that are forbidden by the Pauli principle.  The probability of formation of $^{12}\rm{C}$--$p$  and $^{12}\rm{C}$--$n$ cluster structures in the ground state of $^{13}\rm{N}$ and $^{13}\rm{C}$, respectively, is very high.

Nucleus-nucleon interaction is determined in PCM, like in the shell model, by creating the nuclear field which forms clusterization of nucleus. The potential of cluster wave function is selected in such a way that it describes the cluster scattering phase-shifts taken out from the experimental data. Here in our case of $\rm ^{13}C$ nucleus, we considered clusterization of $A_{1}$=$^{12}\rm{C}$ and  $A_{2}=n$ and for $^{13}\rm{N}$ nucleus  clusterization of $A_{1}$=$^{12}\rm{C}$ and  $A_{2}=p$.

PCM is widely used for the study of nuclei and their interactions owing to its simple nature (see \cite{ebran} and references therein).  It gives the possibility to reduce many-particle problem to two-particle problem. The main problem of PCM is its inability to explain many-particle systems when the internal structure of cluster cannot be ignored. In this account shell model is one of the successful approaches which minimize the many-body problem to one body problem in the self-consistent field of nucleons. There is a peculiar  relation between the cluster model and the shell model and is explained by nucleon association model.
The wave function of cluster nuclei are selected as the ground state wave functions of nuclei $A_{1}$ and $ A_{2}$. So, we assume $A_1=$$^{12}\rm{C}$ and $A_2=p$ ($n$) in the case of  nucleus $^{13}\rm{N}$ ($^{13}\rm{C}$).

The wave function of $^{13}\rm{N}$ ($^{13}\rm{C}$) is considered as a product of the wave function of $^{12}\rm{C}$ (core nucleus) and the wave function of the relative motion of proton (neutron) in the field of $^{12}\rm{C}$ nucleus.

We choose the Woods-Saxon potential, as the potential that determines the field in which the nucleon moves,
\begin{equation}
	V_{N} (r)=-\bigg[V_{0} -V_{LS} (\mathbf{L}\cdot \mathbf{S})\frac{2}{m_{\pi
		}^{2} } \frac{d}{rdr} \bigg]\frac{1}{1+\exp \big(\frac{r-R_{N}}{a}\big)} ,
	\label{W-D pot}
\end{equation}
where $V_0$ ($V_{LS }$) is the depth of the central (spin-orbital)
potential; $\mathbf{L}$ is the angular momentum operator for the
relative motion of the particles; $\mathbf{S}$ is the spin operator;
$m_\pi$ is the pion mass; \textit{a} is the diffuseness and
$R_N=r_0A^{1/3}$ ($r_0$ is the radius parameter of the nuclear
potential; \textit{A} is the atomic mass number). The Coulomb
interaction potential of proton with the $^{12}\rm{C}$ nucleus is taken of the form
\begin{equation}
	\label{Coulomb pot}
	V_C(r)=\cases{\frac{{Z_Ce^2}}{2R_C}\big(3-\frac{r^2}{R_C^2}\big)
		& $r<{R_C}$,\\
		\frac{Z_Ce^2}{r} &$r>{R_C}$	}
\end{equation}
where $Z_{C}e$ is the charge of $^{12}\rm{C}$;
$R_C=r_C\,A^{1/3}$ ($r_C$ is the Coulomb radius parameter). This Coulomb potential corresponds to the case of a uniform charge distribution in the nucleus.

The radial wave function $\varphi_l(r)$ for the partial wave with the orbital momentum \textit{l} is the solution of the radial Schr\"odinger equation ($\mu$ is the reduced mass of nucleon and $^{12}\rm{C}$ nucleus, $E$ is the
energy in center of mass (CM) system)
\begin{equation} \label{Shr-Eq}
	\left\{\frac{d^2}{dr^2} +2\mu
	\left[E-V(r)\right]-\frac{l(l+1)}{r^2}\right\}\varphi_{l} (r)=0,
\end{equation}
where $V(r)$ is sum of the nuclear and the Coulomb potentials for proton motion and the nuclear potential for neutron motion.  The radial wave function $\varphi_l(r)$ satisfies the standard boundary condition at the
origin:
\begin{equation} \label{initial value}
	\left. \varphi_{l} (r)\right|_{r=0} =0.
\end{equation}

Both  $^{13}\rm{N}$ and $^{13}\rm{C}$ nuclei occupy the state $J^\pi=1/2^-$ because this quantum number is determined by the nucleon in the external orbital $1p_{1/2}$.

Accordingly, we take the following parameters of the potential Eq.~(\ref{W-D pot})  for the calculation of the bound state wave function of $^{13}$N$=$$^{12}$C-$p$ (the first pair):
\begin{eqnarray}\label{parameters-NN}
	V_0& =& 45.592 \,\,\rm{ MeV},\nonumber\\
	V_{LS}&=& 7.0\,\,\rm{MeV},\nonumber\\
	a& =& 0.635 \,\,\rm{fm},\nonumber\\
	R_N&=&R_c=1.25\times 12^{1/3}=2.86\,\,\rm{fm}.
\end{eqnarray}
Earlier we used the same potential for calculation of the radiative capture reaction ${^{12}\rm{C}}(p,\gamma){^{13}\rm{N}}$  at low energies  (see Ref. \cite{Kabir}). In a similar fashion  for the calculation of the bound state wave function of $^{13}$C$=$$^{12}$C-$n$  the potential parameters are as listed below
\begin{eqnarray}\label{parameters-NC}
	V_0& =& 45.547 \,\,\rm{ MeV},\nonumber\\
	V_{LS}&=& 7.0\,\,\rm{MeV},\nonumber\\
	a& =& 0.635 \,\,\rm{fm},\nonumber\\
	R_N&=&1.25\times 12^{1/3}=2.86\,\,\rm{fm}.
\end{eqnarray}
Additionally, for the second pair of potentials, we slightly change few parameters as follows:

for $^{13}\rm{N}$ nucleus
\begin{eqnarray}\label{parameters-NN-1}
	V_0& =& 45.543 \,\,\rm{ MeV},\nonumber\\
	V_{LS}&=& 7.0\,\,\rm{MeV},\nonumber\\
	a& =& 0.635 \,\,\rm{fm},\nonumber\\
	R_N&=&1.25\times 12^{1/3}=2.86\,\,\rm{fm}\nonumber\\
	R_c&=&1.28\times 12^{1/3}=2.93\,\,\rm{fm}
\end{eqnarray}

and for $^{13}\rm{C}$ nucleus
\begin{eqnarray}\label{parameters-NC-1}
	V_0& =& 45.536 \,\,\rm{ MeV},\nonumber\\
	V_{LS}&=& 7.0\,\,\rm{MeV},\nonumber\\
	a& =& 0.6 \,\,\rm{fm},\nonumber\\
	R_N&=&1.25\times 12^{1/3}=2.86\,\,\rm{fm}.
\end{eqnarray}
These potentials describe the ground state of the $^{12}$C-$p$ and $^{12}$C-$n$ cluster system.  The overlap integral is 0.997 when using the first pair of potentials and 0.995 for the second pair of potentials.  Denoting the ratio of the squared proton ANC to the neutron one  by $\mathcal R_0$ as given in Ref. \cite{timofeyuk}, we get $ {\mathcal R} _0 = $ 1.169 for the first pair of potentials, and ${\mathcal R}_0=$ 1.254 for the second pair, respectively.  The values of the ratio of the squared ANCs are close to theoretical prediction presented in Table 1 of Ref. \cite{timofeyuk}.
Later in Ref. \cite{okolowicz},  a comparison of the squared ANCs of a one-nucleon virtual separation   was made using calculation of the many-particle Schr\"odinger equation for  several mirror nuclei applying the Gamow shell model and variational Monte Carlo methods. The results, which are given in the article indicate that the squared ANC ratio can vary by up to 30\%. We, therefore take into account this 30\% limit in our present  calculation.

The bound state wave function $\varphi_l(r)$ has the following limits
\begin{eqnarray}\label{definition ANC}
	\varphi_l(r)&\rightarrow&b_lW_{-\eta_b,l+1/2}(2\kappa_br),\,\,\, r\to \infty,\, ({^{12}\rm{C-}}p) \\
	\varphi_l(r)&\rightarrow&b_l\sqrt{\frac{2\kappa_br}{\pi}}K_{l+1/2}(\kappa_br),\, r\to \infty, ({^{12}\rm{C-}}n),
\end{eqnarray}
where $W_{\eta,\nu}(x)$ is the Whittaker function; $K_\nu(x)$ is the modified Bessel function of the second kind; $b_l$ is the single-particle ANC; $\kappa_b=\sqrt{2\mu\varepsilon_b}$, ($\varepsilon_b$ is the binding energy); $\eta_b=Z_ Ce^2\mu/\kappa_b$ is the Sommerfeld parameter. The radial wave function $\varphi_l(r)$  satisfies the normalization condition:
\begin{equation} \label{norm}
	\int\limits_{0}^{\infty}\varphi_l^2(r)dr=1.
\end{equation}
Using a fixed value of the binding energy, it is possible to change the parameters of the potential and obtain different values of the ANC. In other words, there are many potentials that lead to the same binding energy but different ANC values. Taking into account that the ANC is an experimentally derived value from the analysis of scattering  phase-shifts, we will determine this value.

\section{Determination of the ANC from the elastic scattering phase-shift analysis} \label{ANC determination}
In this section, we briefly describe the relationship between the partial amplitude of the elastic scattering and the ANC value of the  bound state wave function tail, given in  Refs. \cite{irgaz2015, orlov2016, orlov2017, irgaz2018}.

The partial amplitude of the nuclear scattering in the presence of the Coulomb interaction is
\begin{equation}\label{amp}
	f_l(k)=\exp(2\rm i\sigma_l) [\exp(2\rm i\delta_l)-1]/2ik,
\end{equation}
where
\begin{equation} \label{Coulomb phase}
	\exp(2\rm i\sigma_l) =
	{\Gamma(l+1+\rm i\eta)}/{\Gamma(l+1-\rm i\eta)}.
\end{equation}
Here $\delta_l$ is the nuclear phase-shift modified by the Coulomb interaction, and $\eta=\xi/k$ is the Sommerfeld parameter, $\xi=Z_1Z_2\mu\alpha$, $\sigma_l$ is a phase-shift of  pure Coulomb scattering, $k=\sqrt{2\mu E}$, $\mu$, $E$  are the relative momentum, the reduced mass and the center-of-mass (cms) energy  of the colliding nuclei with the charge numbers $Z_1$ and $Z_2$,  respectively, and $\alpha$ is the fine-structure constant.

The amplitude Eq.~(\ref{amp}) has a complicated analytical property in the complex momentum plane of $k$ due to the Coulomb factor. According to Ref. \cite{Hamilton}, we can renormalize  the partial amplitude of the elastic scattering multiplying it by the function
(the Coulomb correcting or re-normalizing factor $CF_l(k)$)
\begin{equation} \label{yost}
	CF_l(k)=\frac{(l!)^2e^{\pi\eta}}{(\Gamma(l+1+\rm i\eta))^2}.
\end{equation}
The  renormalized amplitude  has the same analytical  properties  as the amplitude of the scattering of an uncharged particle in the short-range potential field.
The general pole condition $\cot\delta_l-i=0$ follows from the expression for the re-normalized amplitude of the elastic scattering
\begin{equation}\label{fl2}
	\tilde{f}_l(k)=\frac{1}{k(\cot\delta_l-\rm i)\rho_l(k)},
\end{equation}
where the function $\rho_l$ is defined by the equation
\begin{equation}\label{rho}
	\rho_l(k)=\frac{2\pi\eta}{e^{2\pi\eta}-1}\prod_{n=1}^l\Bigl(1+\frac{\eta^2}{n^2}\Bigr).
\end{equation}
Note that in neutron scattering from the nucleus, the factor $\rho_l$ is equal to unity.

The renormalized scattering amplitude of the effective range  method  is written as
\begin{equation}\label{renormalized amplitude}
	\tilde f_l (k) = \frac{k^{2l}}{K_l(k^2)-2\xi D_l(k^2)h(\eta)},
\end{equation}
where $K_l(k^2)$ is given by
\begin{equation} \label{CoulombKl}
	K_l(k^2) = 2 \xi D_l(k^2)\left[C_0^2(\eta)(\cot\delta_l - \rm i)+h(\eta )\right].
\end{equation}
This function is called the effective-range function of the nuclei elastic scattering which is an analytical function on the complex plane of momentum $k$.
Here we denote
\begin{eqnarray}
	C_0^2(\eta)&=&\frac{\pi} {\exp(2\pi\eta)-1},\label{C02}\\
	h(\eta)&=&\psi(\rm i\eta) + \frac{1}{2\rm i\eta} - \ln(\rm i\eta),\label{h-eta}\\
	D_l(k^2)&=&\prod_{n=1}^l\Bigl(k^2+\frac{\xi^2}{n^2}\Bigr),\qquad D_0(k^2)=1,\label{DL_k}
\end{eqnarray}
and $\psi(x)$ is the digamma function. We note that the
effective-range function  $K_l(k^2)$ is real and continuous on the positive real axis of the
energy plane and can be analytically extended from the real positive axis to pole energies situated in the complex plane of the momentum $k$ or energy $E$.
The effective-range function is real for both positive and negative energy values and has no peculiarities at zero, further
\begin{equation}\label{al}
	K_L(0)=-1/a_l,
\end{equation}
where $a_l$ is the scattering length.

For neutron scattering we get from Eq. (\ref{CoulombKl})
\begin{equation} \label{neutronKl}
	K_l(k^2) = k^{2l+1}\cot\delta_l.
\end{equation}
We define the $\Delta_l(k^2)$ function as
\begin{equation}\label{Delta-R}
	\Delta_l(k^2)=C_0^2(\eta)\cot\delta_l,
\end{equation}
on the positive energy semi-axis. So, the renormalized amplitude Eq.~(\ref{renormalized amplitude}) can be re-written as
\begin{equation}\label{fl3}
	\tilde{f}_l(k)=\frac{k^{2l}}{2\xi D_l(k^2)\big(\Delta_l(k^2)-i C_0^2(\eta)\big)},
\end{equation}
where the Coulomb function $h(\eta )$ in Eq. (\ref{fl3}) is absent.

The $\Delta_l(k^2)$ is a function of $k^2$ and it may be expanded as a series (or the Pad\'e expansion  if we take into account the resonant state close to the threshold)   over $k^2$ (or $E$). We can restrict to the first few terms of the expansion at low energy.

The renormalized amplitude continuation from positive to negative energies is given as (see  (13) of Ref. \cite{orlov2017})
\begin{equation}\label{flneg}
	\tilde{f}_l(k)=\frac{k^{2l}}{2\xi D_l(k^2)\Delta_l(k^2)},
\end{equation}
where $k^2$ is negative, and $\Delta_l(k^2)$ function at the negative semi-axis of energy is continuation of the polynomial (or the Pad\'e approximant) of the $\Delta_l$ function. Note that Eq. (\ref{flneg}) follows directly from  the effective-range theory if we use the asymptotic expansions of the digamma function in Eq. (\ref{h-eta}) in low-energy region and substitute this decomposition into Eq. (\ref{renormalized amplitude}).  At the energy $E=-\varepsilon_b$ ($\varepsilon_b$ is the binding energy) the amplitude $\tilde{f}_l(k)$ has pole therefore $\Delta_l(k)=0$ at the same point.

Next, we define how to extend the effective-range function Eq.~(\ref{neutronKl}) to determine the pole of the scattering amplitude for neutron scattering on the nucleus. In this case, using the conventional relation of the scattering amplitude with the effective-range function,
\begin{equation}\label{fl-uncharge}
	f_l(k)=\frac{k^{2l}}{K_l(k^2)-ik^{2l+1}},
\end{equation}
we obtain the condition for the amplitude pole for  negative energies:
\begin{equation}\label{neutron-pole}
	K_l(k^2)- \rm ik^{2l+1}=0,
\end{equation}	
where $k^2<0$.
Finally, calculating the residue of the scattering amplitude  Eq.~(\ref{flneg}) ($W_l=res \tilde{f}_l(k)$) or Eq.~(\ref{fl-uncharge}) ($W_l=res f_l(k)$) depending on the type of scattered particle  (charged or uncharged), we determine the ANC of the wave function of a bound state:
\begin{eqnarray}
	C_l&=&\frac{\sqrt{2\kappa_b}\,\Gamma(l+1+\eta_b)}{l!}|W_l|,\quad \rm{charged\,\, particle}\label{ANC_p}\\
	C_l&=&\sqrt{2\kappa_b}|W_l|,\qquad\qquad\qquad\quad \rm{uncharged\,\, particle}\label{ANC_n}
\end{eqnarray}
where $\kappa_b=\sqrt{2\mu\varepsilon_b}$ and $\eta_b=\xi/\kappa_b$.

\section{Extraction of the ANC{\lowercase{s}} from the experimental phase-shift data and calculation of {\lowercase{log}}$\,{\it ft}$ and half-life values of $^{13}\rm{N}$}\label{extraction}
The nuclei  $^{13}\rm{N}$ and  $^{13}\rm{C}$ are mirror nuclei, therefore the quantum numbers are same ($J^\pi=1/2^-,\,\,T=1/2$) and determined by outermost nucleon in the level $p_{1/2}$ according to the shell model.
The binding energies of  $^{12}{\rm{C}}-p$ and  $^{12}{\rm{C}}-n$ systems are $\varepsilon_b= 1.994$ MeV and $\varepsilon_b= 4.946$ MeV, respectively.

The experimental elastic scattering phase-shift includes  many-body effect of the scattered nuclei. The same may be claimed about the ANC extracted from the experimental data of elastic scattering. For this reason, when we correct the single-particle wave function of the bound state so that it has an experimental ANC value, we actually take into account the influence of many-particle effect.

To determine the ANC we fit the $p_{1/2}$ phase-shift scattering of proton and neutron on $^{12}\rm{C}$ at small energy collision. 
We have applied the phase-shift data given in Refs. \cite{Reich, Drigo} (proton scattering) and Refs. \cite{Weil, Lister} (neutron scattering) in this work.

\begin{figure}[htbp]
	\begin{minipage}[t]{0.5\linewidth}
		\includegraphics[width=\linewidth]{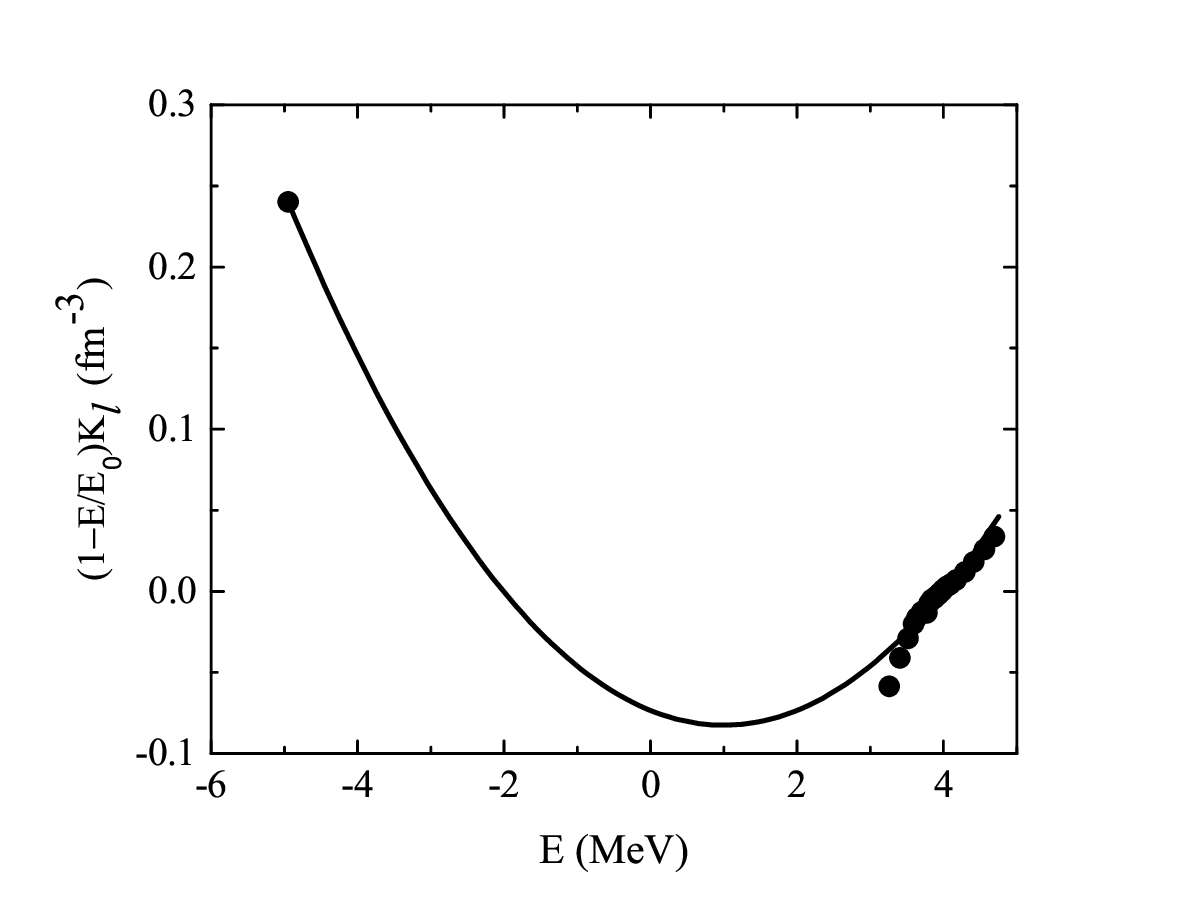}
		\caption{Dependence of our fitted $K_l(E)$ function  (\ref{fitKl}) vs. the
			center-of-mass energy $E$ of the $n$-$^{12}\rm{C}$ collision in  $J^\pi=1/2^-$ state. Solid line is the theoretical calculation. The experimental data (dots) correspond to the phase-shifts taken from Ref. \cite{Weil}.}
		\label{fig1}
	\end{minipage}
	\hspace{0.5cm}
	\begin{minipage}[t]{0.5\linewidth}
		\includegraphics[width=\linewidth]{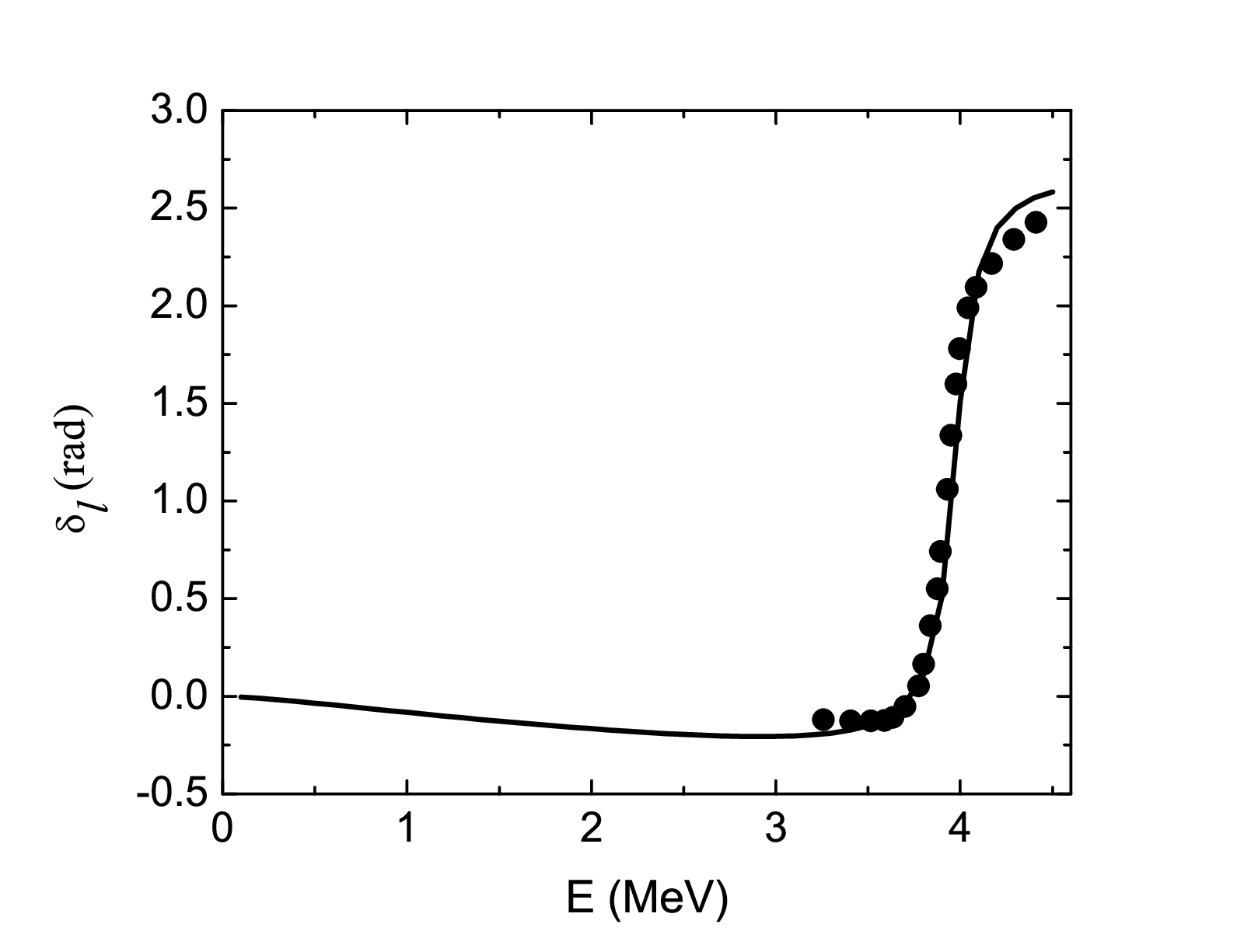}
		\caption{Comparison of the restored phase-shifts (solid line) with the experimental phase-shift data (dots) of $n$-$^{12}\rm{C}$ elastic collision in  $J^\pi=1/2^-$ state. The experimental data are taken from Ref. \cite{Weil}.}
		\label{fig2}
	\end{minipage}
\end{figure}

The experimental phase-shifts of neutron scattering at $J^\pi=1/2^-,\,T=1/2$ cover the energy region including a resonance at energy of $E_x = 8.860 \pm 20$ MeV and a width of $\Gamma = 150 \pm 30$ keV. Therefore, the effective-range function $K_l(k^2)$ [Eq. (\ref{neutronKl})] fits by the Pad\'e approximant  taking into account $E=k^2/2\mu$ is given by
\begin{equation}\label{fitKl}
	K_l(E)=(a_0+a_1E+a_2E^2+a_3E^3)/(1-E/E_0),
\end{equation}
where $E_0$ is the value of energy where the phase-shift becomes zero at the intersection of the positive energy semi-axis. We note that there is a resonance near this point.  This fitting is carried out with the data of the scattering phase-shifts presented in Refs. \cite{Weil, Lister}. It is noted that  we include the point that determines the binding energy to the number of points required for fitting the polynomial Eq.~(\ref{fitKl}).
This is known with greater accuracy than the phase-shifts of scattering. At this point, the effective-range function is equal to
\begin{equation}\label{fitKlE_b}
	K_l(-\varepsilon_b)=(2\mu\varepsilon_b)^{3/2}\qquad (l=1).
\end{equation}
Since part of the elastic scattering phase-shifts was measured at energies above the excitation level of $^{12}\rm{C} (2^+)$, we exclude the phase-shifts  above the excitation level ($E_x=4.43982\,\,\rm{MeV} \pm 0.21$ keV) from fitting procedure. We also  exclude points close to zero energy, so that the ratio of the squared ANCs  does not exceed $\sim$30\% of the limit, as mentioned before.
The results of fitting the data from Ref. \cite{Weil} and the graph of the restored scattering phase-shifts (using the Pad\'e approximant) are shown in Figures \ref{fig1} and ~\ref{fig2}, respectively. To avoid the singularity of the fit approximant equation (\ref{fitKl}) at the point $E_0$, we multiply it by the factor ($1-E/E_0$), as can be seen in Figure \ref{fig1}.  We also calculate the resonance position and its width, and got the following results:  $E_r=3.959$ MeV, $\Gamma=174$ keV. These results are close  to the data presented in Ref. \cite{nucl-data} ($J^\pi=1/2^-,\,\,E_x=8.860\pm 0.02\,\rm{MeV},\,\, \Gamma=150\pm 30\,\rm{keV}$).

The same fitting procedure is carried out with the experimental data from Ref. \cite{Lister}.

\begin{figure}[htbp]
	\begin{minipage}[t]{0.5\linewidth}
		\includegraphics[width=\linewidth]{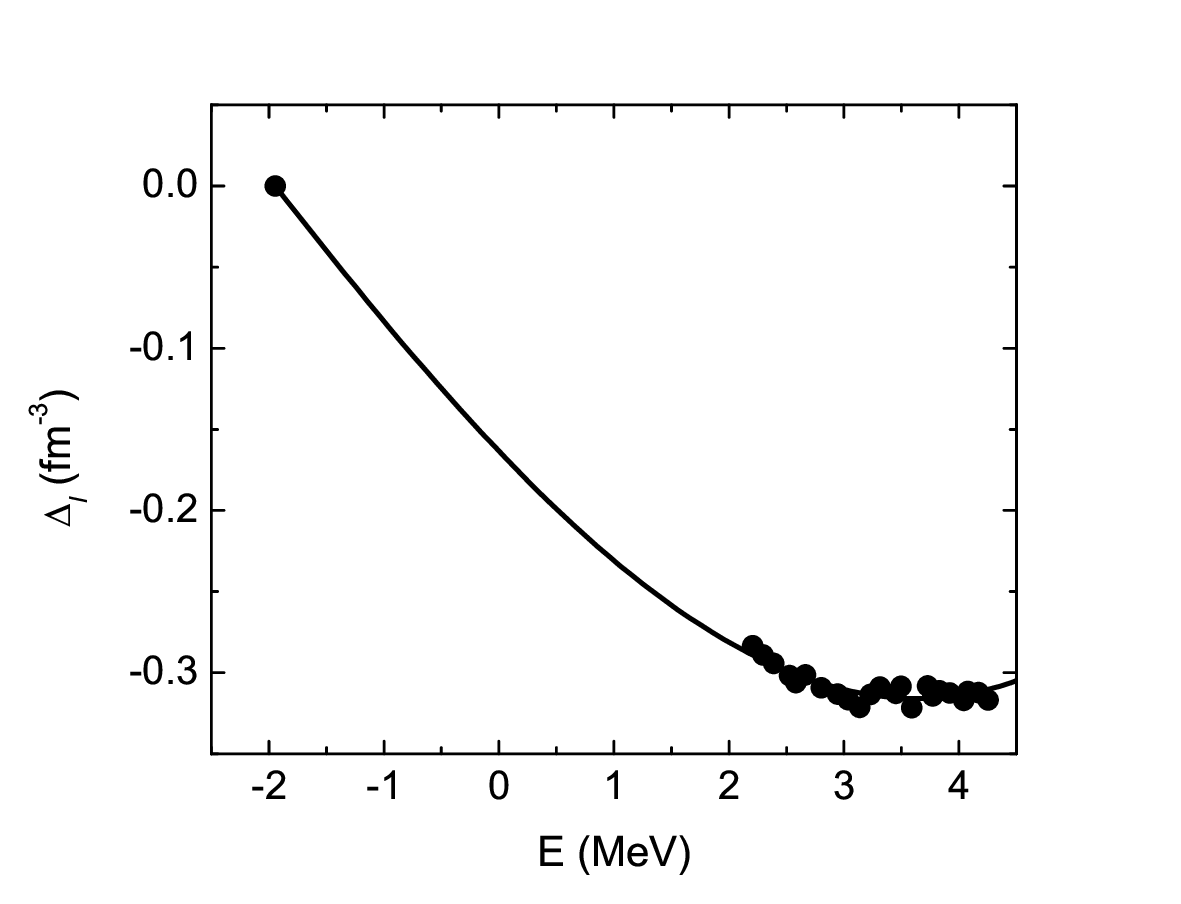}
		\caption{Dependence of our fitted $\Delta_l(E)$ function  Eq.~(\ref{fitDeltalE}) vs. the center-of-mass energy $E$ of the $p$-$^{12}\rm{C}$ collision in $J^\pi=1/2^-$ state. Solid line is the theoretical calculation. The experimental data (dots) correspond to the phase-shifts taken from Ref. \cite{Reich}}
		\label{fig3}
	\end{minipage}
	\hspace{0.5cm}
	\begin{minipage}[t]{0.5\linewidth}
		\includegraphics[width=\linewidth]{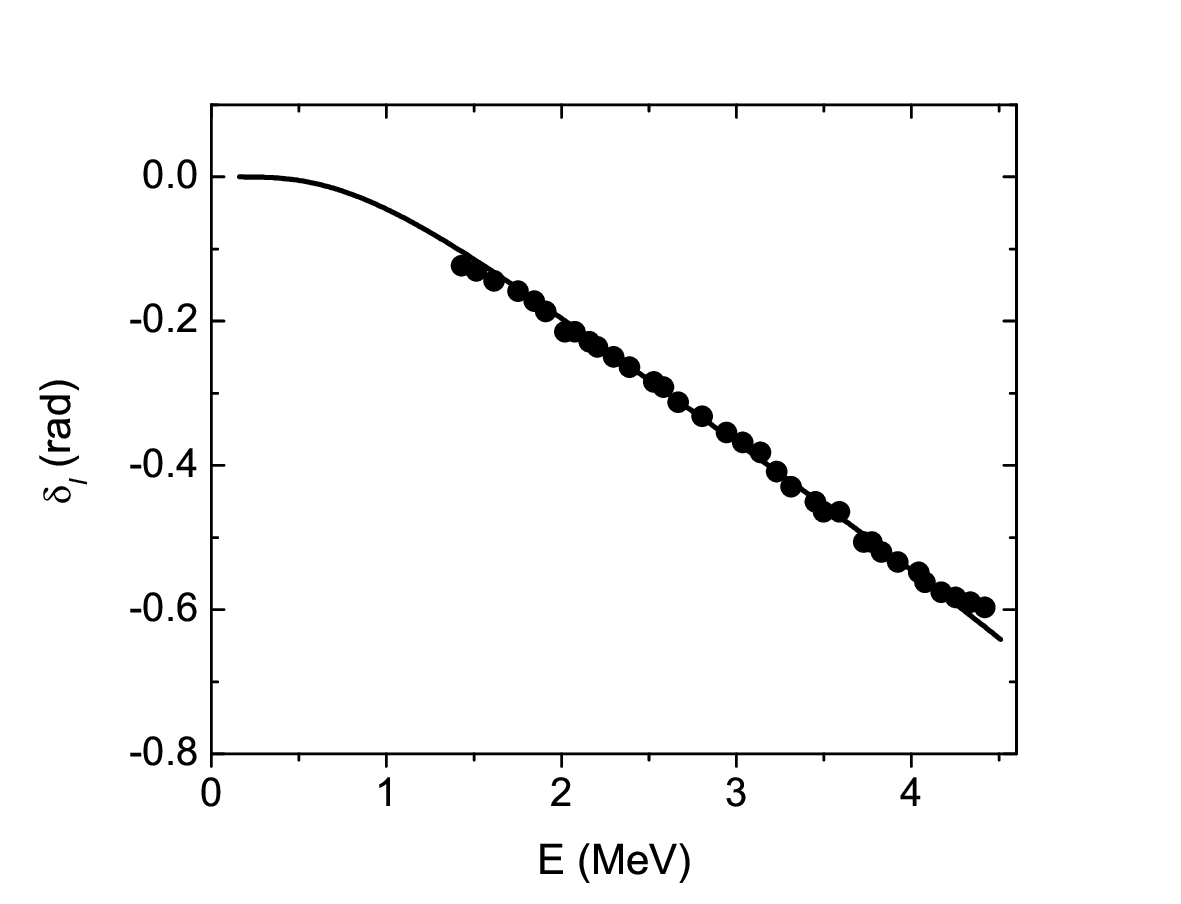}
		\caption{Comparison of the restored phase-shifts (solid line) with the experimental phase-shift data (dots) of $p$-$^{12}\rm{C}$ elastic collision in  $J^\pi=1/2^-$ state. The experimental data are taken from Ref. \cite{Reich}.}
		\label{fig4}
	\end{minipage}
\end{figure}

\begin{figure}[htbp]
	\begin{minipage}[t]{0.5\linewidth}
		\includegraphics[width=\linewidth]{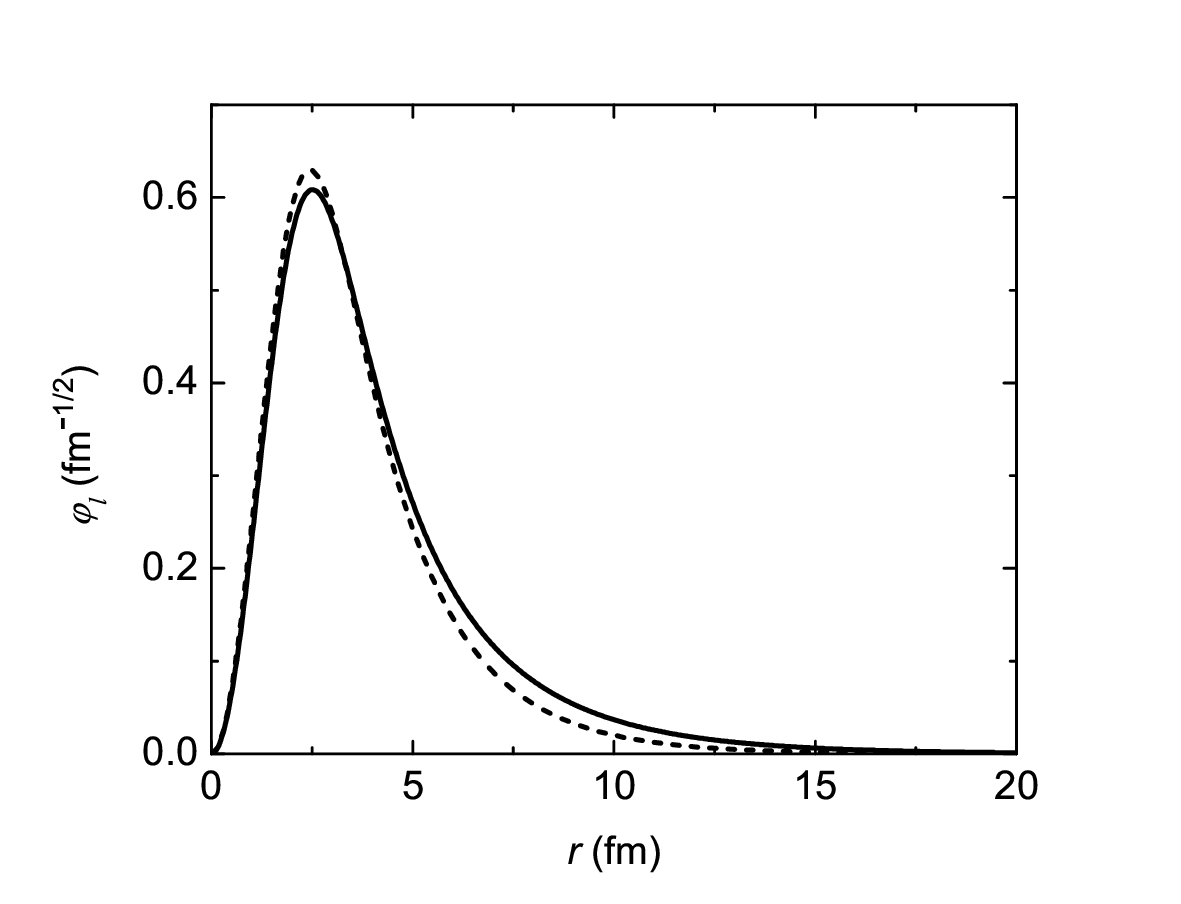}
		\caption{Comparison of the $n$-$^{12}\rm{C}$ (dashed line) and $p$-$^{12}\rm{C}$ (solid line) wave functions obtained from solving the Schrödinger equation with the parameters of the potentials given in Eqs. (\ref{parameters-NN}) and (\ref{parameters-NC}).}
		\label{fig5}
	\end{minipage}
	\hspace{0.5cm}
	\begin{minipage}[t]{0.5\linewidth}
		\includegraphics[width=\linewidth]{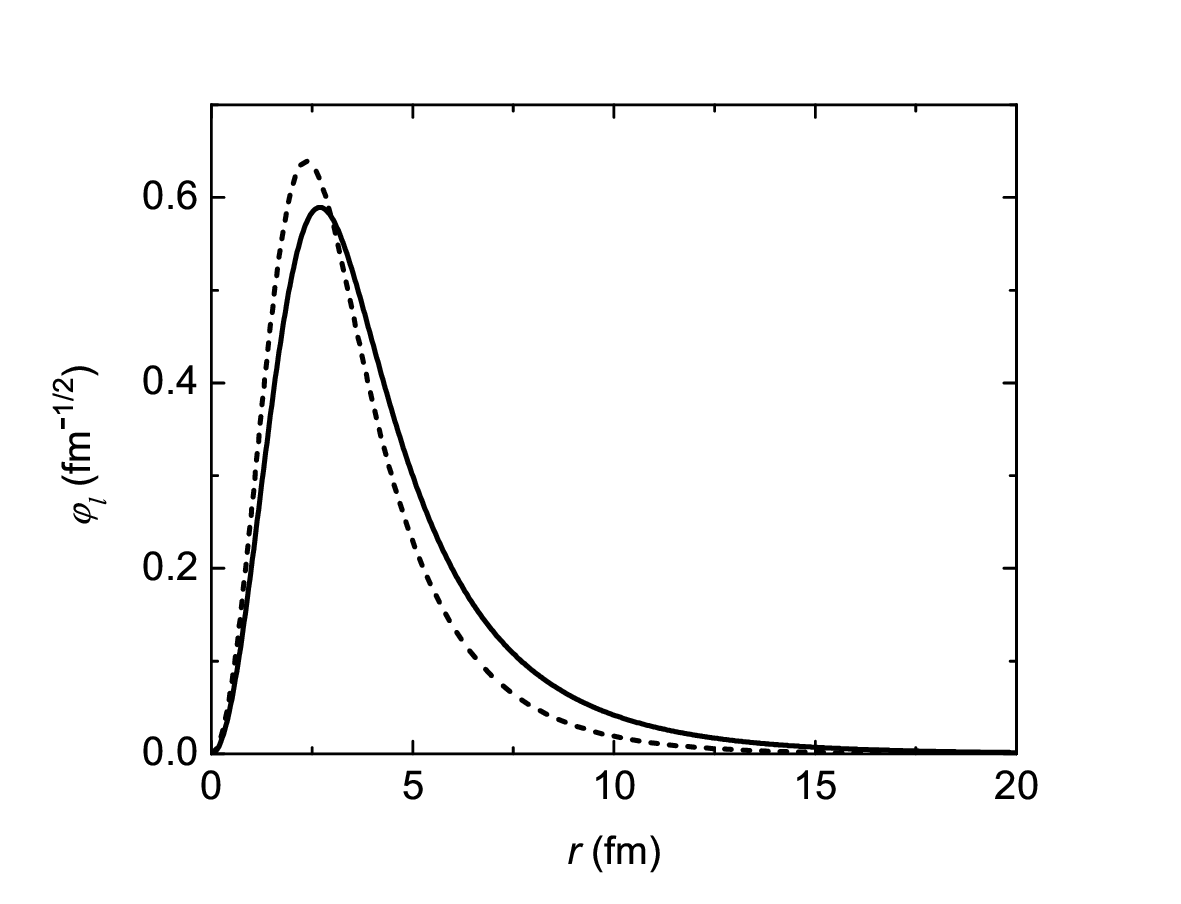}
		\caption{Comparison of the $n$-$^{12}\rm{C}$ (dashed line) and $p$-$^{12}\rm{C}$ (solid line) wave functions obtained using the inverse scattering theory. The single-particle ANCs equal to the  experimental ANC values ($C_n=$1.694 fm$^{-1/2}$ and $C_p=$2.216 fm$^{-1/2}$).}
		\label{fig6}
	\end{minipage}
\end{figure}

The fitting procedure in the case of $p$-$^{12}\rm{C}$ is determined using the $\Delta$-method is described previously in Refs. \cite{orlov2017, irgaz2018}. The experimental data of phase-shifts, suitable for fitting, are known up to 5.0 MeV, which is less than the resonance energy in the $p_{1/2}$ state ($J^\pi=1/2^-$, $E_x=8.918\pm 0.011$ MeV, $\Gamma=230$ keV). Therefore, for the fitting procedure we use the polynomial
\begin{equation}\label{fitDeltalE}
	\Delta_l(E)=(1+E/\varepsilon_b)(a_0+a_1 E+a_2 E^2).
\end{equation}
Here we use the factor $(1+E/\varepsilon_b)$, because the $\Delta$-function must be zero at the pole point of the scattering amplitude corresponding to the $p$-$^{12}\rm{C}$ binding energy ($\varepsilon_b=1.944$ MeV). The results of fitting the data from Ref. \cite{Reich} and the graph of the restored scattering phase-shifts using  the polynomial  are shown in Figures \ref{fig3} and \ref{fig4}, respectively.
We used the experimental data from Ref. \cite{Drigo}  for fitting as well.
From the Figures, we note that fitting with functions  (\ref{fitKl}) and (\ref{fitDeltalE}) allows us to describe the phase-shifts of elastic scattering in the low-energy region rather well. However, the obtained ANC values for $n$-$^{12}\rm{C}$ and $p$-$^{12}\rm{C}$ systems using these functions differ in values calculated from the solution of the Schr\"odinger equation (\ref{Shr-Eq}) (see Table 1). As mentioned earlier, the Woods-Saxon potential can be modified in such a way that solving the Schr\"odinger equation with modified potentials may leads to the desired ANC value. Indeed, using the theorem of the inverse problem in scattering theory Ref. \cite{Chadan-77} we obtain the modified phase equivalent potential leading to the experimental  ANC value. We adopt this nuclear potential $V_N(r)$, which we apply to calculate  the bound-state wave function. The phase equivalent potential is given by
\begin{equation}
	V^{(1)}_{N}(r)=  V_{N}(r) + \frac{d^2K(r)}{dr^2},
	\label{vn11}
\end{equation}
\begin{equation}
	K(r)=   {\rm ln}[1 + (\tau - 1)(1- \int\limits_{0}^{r}\, dx\varphi_l^{2}(x))].
	\label{kr1}
\end{equation}
The new bound state wave function in the potential  $V^{(1)}_{N}(r)$ can be expressed in terms of the bound state wave function $\varphi_l(r)$ as:
\begin{equation}
	\varphi_l^{(1)}(r)= \tau^{1/2}\,\frac{\varphi_l(r)}{1 + (\tau -1)\, \int\limits_{0}^{r}\, dx\varphi_l^{2}(x)}
	\label{varphi11}
\end{equation}
where $\varphi_{l}(r)$ is the solution of Eq. (\ref{Shr-Eq}).  By varying the parameter $\tau$  we get the required ANC value.
Using the procedure described above we find a new wave function whose tail contains the fitted value of ANC instead of the single-particle ANC.   Figures \ref{fig5} and \ref{fig6} show the comparison of the wave functions $n$-$^{12}\rm{C}$ and $p$-$^{12}\rm{C}$, before and after the change of the wave functions according to the inverse scattering theory. One notes that the values of the wave function $p$-$^{12}\rm{C}$ changed slightly both in the external and internal regions in comparison to the original single-particle function.
The result of the calculation of the single-particle and fitted  ANC values, squared overlap integrals, $\log\rm{\textit{ft}}$ and half-life values are shown  in Table 1.
\begin{table}[t]
	\caption{\label{tab1}The first and second column represents the ANC for  $\rm ^{13}N$ and  $\rm ^{13}C$, respectively. The third, fourth and fifth columns represent the squared overlap integral, $\log\rm{\textit{ft}}$ value and half-life of $\rm ^{13}N(\beta^+)^{13}C$ decay, respectively. The experimental values: half-value $T_{1/2}=9.965\pm{0.0004}$ min; log\textit{ft}=3.667$\pm 0.001$ (see Ref. \cite{nucl-data}).}
	\begin{indented}
		\item[]\begin{tabular}{@{}ccccc}
			\br
			$C_p$ ($\rm fm^{-1/2}$)  & $C_n$ ($\rm fm^{-1/2}$)  & $|\langle \psi_f|\psi_i\rangle|^2$   & log\textit{ft} & $ T_{1/2}$ (min.)   \\\hline
			1.951{$^{\rm{a}}$}  & 1.806{$^{\rm{a}}$}  &0.9943& 3.612 & 8.859\\
			1.953$^{\rm{b}}$  & 1.744{$^{\rm{b}}$}  &0.9918& 3.614 & 8.881\\
			2.162 \cite{Drigo} & 1.712\, \cite{Weil} &0.9756& 3.620 & 9.028\\
			2.216 \cite{Reich} & 1.712 \cite{Weil} &0.9708& 3.623 & 9.073\\
			2.162 \cite{Drigo} & 1.694 \cite{Lister}&0.9739& 3.622 & 9.045\\
			2.216 \cite{Reich} & 1.694 \cite{Lister}&0.9689& $3.624$ &9.091\\
			1.948$^{\rm{c}}$\cite{Reich} & 1.694 \cite{Lister}&0.9890& $3.615$ &8.906\\
			\br
		\end{tabular}
		\item\noindent$^{a}${the single particle ANC obtained from the solution of the Schr\"odinger equations using  parameters given by Eqs. (\ref{parameters-NN}) and (\ref{parameters-NC})}
		\item\noindent$^{b}${the single particle ANC obtained from the solution of the Schr\"odinger equations using parameters given by Eqs. (\ref{parameters-NN-1}) and (\ref{parameters-NC-1})}
		\item\noindent$^{c}${c.m.energy interval of fitting from 2.52 MeV $< E_{cm} <$ 4.17 MeV} \cite{Reich}
	\end{indented}	
\end{table}

The first and second  rows of Table \ref{tab1} show results obtained from solution of the Schr\"odinger equation (\ref{Shr-Eq}) with the nuclear potential parameters defined  by the first pair of Eqs. (\ref{parameters-NN}) and (\ref{parameters-NC}) and the second pair shown in Eqs. (\ref{parameters-NN-1}) and  (\ref{parameters-NC-1}), respectively. The other rows show results
obtained by using Eqs. (\ref{vn11}) and (\ref{varphi11}) to find the modified  bound wave functions $\varphi_l^{(1)}(r)$ whose tails have experimental values of ANCs.  From the results shown in Table \ref{tab1} we see a correlation between ANC and the calculated value of the overlap integral
$|\langle\psi_f|\psi_i\rangle|^2$  leading to a slight variation in log\textit{ft} and half-life values. If we choose a value of $\mathcal R_0=$1.3 for the squared ratio of ANCs as the reference point, then the deviations for the squared ratio of ANCs represented in Table \ref{tab1} below the second-row  lie within  the 2-32\% range relative to the adopted reference point. At the same time, deviations of the calculated half-lives differ from each other by less than 2.5\%. 
From the calculations performed by using single-particle wave functions normalized to unity, we conclude that it is impossible to obtain results close to the experimental value of half-life. That is, it is necessary to take into account the correction of many-particle effects.

In the next section, we describe a new simple method that uses the spectroscopic factors to  take into account the many-particle effects and  calculate the values of the log{\textit ft} and   half-life  of weak $^{13}$N nucleus decay. The spectroscopic factor gives possibility most easily to take into account corrections of many-particle effects.

\section{New method of calculating nuclear matrix elements and results} \label{new method}

We again consider the transition amplitude Eq.~(\ref{Vif}) determining weak decay of  $^{13}$N nucleus. As we noted above, the single-particle shell model describes rather well the low-lying states of the $^{13}$C and $^{13}$N mirror nuclei, considering them to consist of a $^{12}$C core and nucleon on $1p_{1/2}$ orbital. In accordance with Ref. \cite{nucl-data} the experimental values of half-value $T_{1/2}$ and  log\textit{ft} equal  $9.965\pm{0.0004}$ min and $3.667\pm 0.001$, respectively for the $\beta^{+}$ decay of  $^{13}$N. The use of the single-particle wave functions led to the calculated half-life value of $\sim$11\% smaller  than the experimental value.
Further, the values of the magnetic  moments  calculated according to single-particle shell model differ from the measured data by $\sim$10\%, according to our estimates.  And this also testifies in favor of the single-particle state of $^{13}$C and $^{13}$N  nuclei. The results of calculation based on the many-particle model of the ground state of the $^{13}$C  (see Refs. \cite{sinhgam1986, amos1989}) also confirm the significant contribution of the single-particle $^{12}$C--n configuration to the many-particle wave function of  $^{13}$C nucleus.

The following assumptions are made in presenting a new approximate method for calculation of the nuclear matrix element  Eq.~(\ref{Vif}): 1) the probability of a single-particle shell description of nuclei is rather high. 2) the transition operator acts only on the proton located at the $p_{1/2}$ level and does not act on the nucleons located on the $s_{1/2}$ and $p_{3/2}$ shells.
We insert the identity operator
	\begin{equation} \label{unit operator}
		\hat I=\sum_\alpha|\Phi_\alpha\rangle \langle \Phi_\alpha|,
	\end{equation}
between the many-particle wave function $\Psi_f$ of $^{13}$C nucleus  and the operator $V$  of  Eq. (\ref{Vif}). In Eq. (\ref{unit operator})   $\Phi_\alpha$ are a set of eigenfunctions of the many-particle Hamiltonian that describe all kinds of states of 12 interacting nucleons (6 protons and 6 neutrons) including the ground, excited states of the $^{12}$C nucleus, as well as states with continuous energy. Summation is carried out over all quantum numbers, including isospin. Taking into account that the ground wave function of $^{13}$C ($^{13}$N) includes mainly the $^{12}$C--$n$ ($^{13}$N--$p$) configuration, we obtain the equation
\begin{equation} \label{Vif-1}
	M_{fi}=\langle I_n|V| I_p \rangle,
\end{equation}
where $I_n (({\mathbf r}),\,\,(I_p(({\mathbf r}))$ is the overlap function of the bound state function of  $^{13}$C ($^{13}$N) and bound state function of $^{12}$C (we omit the corresponding quantum numbers for simplicity). Our calculation lead us finally to  the following equation instead of Eq. (\ref{Mifav}):
\begin{equation}\label{Mifav-1}
	\overline{|{\cal{M}}_{fi}|^2}={\langle{\mathcal I}_n|{\mathcal I}_p \rangle}^2(g_{\rm V}^2+g_{\rm A}^2/3),
\end{equation}
where ${\mathcal I}_n$ and ${\mathcal I}_p$ are the radial overlap functions. Following results of Ref. \cite{akram1999} (see also the references therein) we can write relation between the radial overlap function and the single-particle wave function as
\begin{equation}\label{rad-overlap}
	{\mathcal I}_l(r)=S_l^{1/2}u_l(r),
\end{equation}
where $S_l$ is the spectroscopic factor.
The spectroscopic factor is conventionally employed as a measure of the purity of the single-particle configuration of the initial (final) bound wave function of the nucleus. From Eq. (\ref{rad-overlap}) we note that
\begin{equation}\label{norm_S}
	S_l=\int\limits_0^\infty{\mathcal I}_l^2(r)dr,
\end{equation}
and the relationship of the ANC ($C_l$), defined by Eq. (\ref{ANC_p}) [Eq. (\ref{ANC_n})] with the single-particle ANC ($b_l$), is expressed by the relation:
\begin{equation}\label{relation_C-b}
	C_l=S_l^{1/2}b_l.
\end{equation}
Since ANC is an unambiguous value extracted from experiments, akin to the energy level and spin of nucleus, it is a model-independent quantity (in contrast to the spectroscopic factor).
The value of the extracted spectroscopic factor strongly depends on the model used, the type of reaction and the collision energy. For this reason it has a very large range of values (see Refs. \cite{li-2009,liu-2004,tsang-2005} and references therein). The reader may see the spectroscopic factors extracted by different experiments and theoretical calculations listed in Table 2 of Ref. \cite{li-2009}. These values lie in a wide range (from a value of 0.25 to 1.48). We further note the comments from the authors of Ref. \cite{mukham2010}  that in some cases the exact reaction amplitudes could not be parametrized in terms of the spectroscopic factors. We note that the ANC value for $^{13}{\rm{N}}\rightarrow {^{12}}{\rm{C}}+p$ extracted from different experimental results  lies in the interval 1.44-1.82 fm$^{-1/2}$ (see Refs. \cite{li-2009, tashkent2006, nunes2000}). In Ref. \cite{nunes2000} the extracted squared ANC was $3.31\pm 0.45$ fm$^{-1}$, which is close to the value of the single-particle ANC. We also note the article \cite{huang2010} using the experimental ANC value of 2.05 fm$^{-1/2}$ when studying the radiative capture $^{12}{\rm{C}}(p,\gamma)^{13}\rm{N}$. 

The results from the $^{13}\rm{C}\rightarrow {^{12}}\rm{C}+n$ extracted ANC  are presented in Refs. \cite{Goncharov, Belyaeva, li-2001}, The minimum value of the neutron ANC is 1.493 fm$^{-1/2}$ (Ref. \cite{Belyaeva}), while its maximum value is  1.93 fm$^{-1/2}$  (Ref. \cite{li-2001}). Note that in order to obtain the listed results of both proton and neutron ANCs, different values of the spectroscopic factor were used.

Finally, we write the equations for log\textit{ft} and half-life $T_{1/2}$:
\begin{eqnarray}
	T_{1/2}&=&\frac{\kappa}{f_0 S_n S_p\langle\varphi_f|\varphi_i\rangle^2(g_{\rm V}^2+g_{\rm A}^2/3)},\label{half-life-1}\\
	{\rm log}ft&=&{\rm log}\frac{\kappa}{ S_n S_p\langle\varphi_f|\varphi_i\rangle^2(g_{\rm V}^2+g_{\rm A}^2/3)}.\label{logft1}
\end{eqnarray}
From these equations we conclude that in order  for the half-life of $^{13}$N to be bigger than the results shown in the first and second rows of Table \ref{tab1}, the product of the spectroscopic factors $S_n$ and $S_p$ should be less than one, and the deviation of the squared ANCs  ratio should not exceed 30\% of the condition ${\mathcal R}_0=$1.3 which we set according to the result of the articles Refs. \cite{timofeyuk,okolowicz}.

\begin{figure}[thb]
	\begin{center}
		\parbox{12 cm}{\includegraphics[width=12 cm]{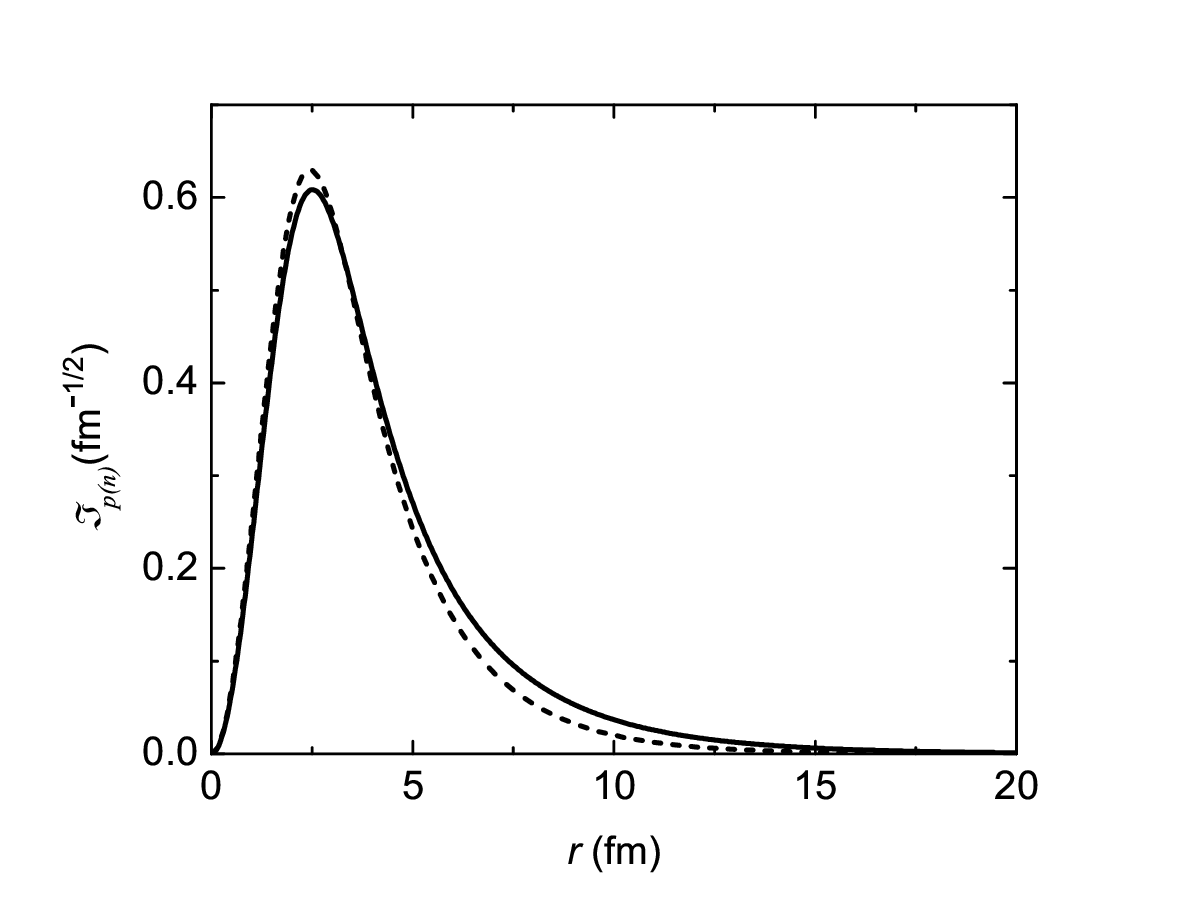}}
	\end{center}
	\caption{Comparison of the $n$-$^{12}\rm{C}$ overlap function ${\mathcal I}_n(r)$ (dashed line) with $p$-$^{12}\rm{C}$ overlap function ${\mathcal I}_p(r)$ (solid line)  having the single-particle ANCs multiplied by roots of the corresponding  spectroscopic factors to get the experimental ANC values in the tails [$b_n(n$-${^{12}\rm{C}})=1.806$ fm$^{-1/2}$ and $b_p(p$-${^{12}\rm{C}})=1.951$ fm$^{-1/2}$;  $S_n=0.880$  and $S_p=0.997$].
		\label{fig7}}
\end{figure}

The nucleon distribution density in the nucleus, starting from $A>10$, is described quite well by the same function as the function describing the Woods-Saxon potential. For this reason, the single-particle wave function $\varphi_l(r)$ and the overlap integral ${\mathcal I}_{n(p)}$ describe the motion of a nucleon inside and outside the nucleus with reasonably  good accuracy. Figure \ref{fig7} presents a comparison of the ${\mathcal I}_n$ and ${\mathcal I}_p$ overlap integrals.
From Figure \ref{fig7}  we see that they are almost the same as in Figure \ref{fig5}, and the curves ${\mathcal I}_n$ and ${\mathcal I}_p$ are close to each other, especially in the inner region of $^{13}$N and $^{13}$C which should be observed for mirror nuclei.  However, the tail of these functions have experimental values of ANCs indicated in the last row of Table \ref{tab1}.

For further calculations, we chose the experimental ANC values obtained by the fitting procedure for the scattering phases and presented in the last row of Table \ref{tab1}, since the probability of a single-particle configuration of the parent ($^{13}$N) and daughter ($^{13}$C) nuclei cannot exceed unity.
Our calculation gave the following results if we apply the data of the single-particle ANCs shown in the first row of Table \ref{tab1}:
\begin{equation}
	S_p=0.997,\,   S_n=0.880,\,  {\rm{log}}ft=3.669,\, T_{1/2}=10.10\,\rm{min}.
\end{equation}
Applying data of the single-particle ANCs shown in  the second
rows of Table \ref{tab1} we get
\begin{equation}
	S_p=0.995,\,   S_n=0.943,\,  {\rm{log}}ft=3.641,\, T_{1/2}=9.461\,\rm{min}.
\end{equation}
Hence we can state the fact that taking into account spectroscopic factors calculated from the ratio of the experimentally measured ANC to the single-particle ANC allows one to correctly calculate the values of log{\textit ft} and half-live   without resorting to complex calculations of the wave functions of multi-particle nuclei.

\section{Summary}\label{summary}
\begin{enumerate}
	\item
	In this work we present ANCs and spectroscopic factors as one of the crucial inputs  when we calculate the nuclear matrix element  to determine the half-life and log\textit{ft} values of $\beta^+$ decay process. This is demonstrated in our work using the $\rm{^{13}N(\beta^+)^{13}C}$ decay, where results are  dependent on the ANCs ratio   corresponding to the bound state functions of \textit{p}-$^{12}$C and \textit{n}-$^{12}$C systems.
	\item
	Using the effective range and $\Delta$-methods, we  determine the ANC values using the process of fitting experimental phase-shift data.  The difference between the extracted maximum and minimum values of the ANC is $\sim$13\% for  $^{13}$N, while  $\sim$1.1\% for $^{13}$C. It is desirable to perform new measurements of the elastic scattering phase-shifts of the proton and neutron by $^{12}$C in the low-energy region below the inelastic scattering region as one have to be sure that the extracted values of the ANCs are correct. 
	\item
	$^{13}$N and $^{13}$C  are mirror nuclei and their single-particle wave functions must be  close to each other, especially  in the inner region of the nuclei. This is endorsed by the result of our calculation.
	There is a certain limitation for the ratio of the squared values of the proton ANC to the neutron one.
	The inverse scattering theory allows us to obtain the modified single-particle wave functions  with experimental ANC values. However, these modified single-particle wave functions normalized to unity allows reproducing the experimental data for the  $^{13}$N decay with an accuracy of $\sim$90\%.  The use of overlap functions, normalized  to the corresponding spectroscopic factors, made  it possible to reproduce the experimental nuclear decay data of  $^{13}$N   closer to the experimental data.
	\item In future we plan to carry out a similar calculation to describe the beta decay of the $^{17}$F nucleus which also represents a nucleus with one proton in the outer shell.
\end{enumerate}

\section*{Acknowledgment}
B.F. Irgaziev is thankful to the Ministry of Innovative Development of Uzbekistan for support through  Grant No. OT-F-2-010 and L. D. Blokhintsev for useful discussion.  J.-U. Nabi would like to acknowledge the support of the Higher
Education Commission Pakistan
through project numbers 5557/KPK/NRPU/R$\&$D/HEC/2016, 9-5(Ph-1-MG-7)/PAK-TURK/R$\&$D/HEC/2017 and Pakistan Science Foundation through project
number PSF-TUBITAK/KP-GIKI (02).


\end{document}